\documentclass[aps,prc,reprint,showpacs,superscriptaddress]{revtex4-1}

\usepackage{graphicx}
\usepackage{dcolumn}
\usepackage{bm}
\usepackage{amsmath}

\begin{document}

\title{Nucleon exchange in heavy-ion collisions within stochastic mean-field approach}

\author{B. Yilmaz}
\affiliation{Physics Department, Faculty of Sciences, Ankara University, 06100, Ankara, Turkey}
\author{S. Ayik}
\affiliation{Physics Department, Tennessee Technological University, Cookeville, Tennessee 38505, US}
\author{D. Lacroix}
\affiliation{IN2P3-CNRS, Universite Paris-Sud, F-91406 Orsay Cedex, France}
\author{O. Yilmaz}
\affiliation{Physics Department, Middle East Technical University, 06531, Ankara, Turkey}
\date{\today}

\begin{abstract}
Nucleon exchange mechanism is investigated in deep-inelastic symmetric heavy-ion collisions in the basis of the Stochastic Mean-Field approach. By extending the previous work to off-central collisions, analytical expression is deduced for diffusion coefficient of nucleon exchange mechanism. Numerical calculations are carried out for ${}^{40}$Ca + ${}^{40}$Ca and ${}^{90}$Zr + ${}^{90}$Zr systems and the results are compared with the phenomenological nucleon exchange model. Also, calculations are compared with the available experimental results of deep-inelastic collisions between calcium nuclei. 
\end{abstract}


\maketitle

\section{Introduction}
In the standard mean-field approximation, the many-body wave function is taken as a single Slater determinant constructed from time-dependent single-particle wave functions. These wave functions are determined by time-dependent Hartree-Fock (TDHF) equations, in which the self-consistent mean-field Hamiltonian usually is expressed in terms of a Skyrme-type effective interaction. At low energies, typically at available energy per particle below the nucleon binding energy, the Pauli blocking severely inhibits binary collisions due to short range correlations. As a result, the standard mean-field approximation provides a good description for the average behavior of the collision dynamics \cite{Koonin,Goeke,Simenel,Negele,Davies}. On the other hand, collective motion is treated in nearly classical approximation and fluctuations of collective variables are severely underestimated. For example, the mean-field predictions for the widths of the fragment mass distributions are in general an order of magnitude smaller than the experimental observations.

Much work has been carried out to improve the standard mean-field approximation by incorporating dynamics of density fluctuations into the description \cite{Randrup,Abe,Lacroix1,Lacroix2}. In recently proposed stochastic mean-field approach (SMF), the effect of quantal and thermal fluctuations in the initial state is incorporated into the description in a stochastic manner \cite{Ayik1}. A number of demonstrations provide rather strong support for the validity of the SMF approach. In one of these demonstrations, it is illustrated that in the limit of small amplitude fluctuations, the SMF approach gives rise to the same formula as the one derived by Balian and Veneroni for dispersion of  one-body observables \cite{Balian,Simenel2}. By an adiabatic projection procedure, from the SMF approach it is possible to derive effective equations for slow collective variables. These equations appear as generalized Langevin description in which dissipation and fluctuation forces are connected by the quantal dissipation-fluctuation relation \cite{Mori,Gardiner,Weiss}. In this manner, connection to the Mori formalism is established. For deep-inelastic collisions, by a geometric projection procedure it is possible to deduce transport coefficients associated with macroscopic variables. These transport coefficients have similar form with those that are familiar from phenomenological nucleon exchange model \cite{Randrup2}, but provide a more refined description for transport mechanism. In a recent application, the SMF approach is tested with Lipkin-Meshkov-Glick model, and it is found that the gross properties of the model are well produced in the SMF approach \cite{Lacroix3}. The SMF approach was also extended by including pairing interaction \cite{Lacroix4}.

Recently we investigated nucleon exchange mechanism in the basis of the SMF approach. In the first applications, central collisions of symmetric and asymmetric ions are considered at energies below the Coulomb barrier \cite{Washiyama,Yilmaz}. At these energies fusion does not take place, colliding ions exchange a few nucleons and separate again. We calculated nucleon drift and diffusion coefficients in the semi-classical approximation and in the Markovian limit by ignoring memory effects. However the most interesting cases naturally are investigations of collisions with finite impact parameters because then detailed comparison with experiments can be made. In this work, we investigate nucleon exchange mechanism in off-central deep-inelastic collisions of symmetric heavy-ions. We carry out these investigations also in the semi-classical limit by ignoring memory effects. In section 2, we present a short review of the SMF approach. In section 3, after a discussion of the window dynamics, we present derivation of nucleon diffusion coefficient and illustrate several results of calculations for ${}^{40}$Ca + ${}^{40}$Ca and ${}^{90}$Zr + ${}^{90}$Zr collisions. Also, we compare the results of calculations with the available data for symmetric collisions of calcium nuclei. Conclusions are given in the section 4.

\section{Stochastic Mean-Field Approach}
In order to describe the full quantal evolution and dynamics of fluctuations, in principle, we need to consider a superposition of a complete set of Slater determinants with time dependent expansion coefficients. Unfortunately, this is a very difficult task to carry out. In the SMF approach, rather than considering this large set of Slater determinants, an ensemble of single-particle density matrices is constructed by incorporating quantal and thermal fluctuations in the initial state \cite{Ayik1,Ayik2}. We can express a member of the ensemble (indicated by the event label $\lambda$) as,
\begin{equation} \label{eq1} 
\rho ^{\lambda } (\vec{r},\vec{r'},t)=\sum _{ij}\phi _{i}^{*}  (\vec{r},t;\lambda )\rho _{ij}^{\lambda } \phi _{j} (\vec{r'},t;\lambda ) 
\end{equation} 
where $(i,j)$ indicates a set of quantum numbers specifying the single particle wave functions including spin and isospin degrees of freedom. Time independent expansion coefficients $\rho _{ij}^{\lambda}$ are taken as uncorrelated random numbers specified by Gaussian distributions. The mean value of each Gaussian is determined by,
\begin{equation} \label{eq2} 
\overline{\rho _{ij}^{\lambda } }=\delta _{ij} n_{j}  
\end{equation} 
and its variance is specified according to,
\begin{equation} \label{eq3} 
\overline{\delta \rho _{ij}^{\lambda } \delta \rho _{j'i'}^{\lambda } }=\frac{1}{2} \delta _{ii'} \delta _{jj'} \left[n_{j} (1-n_{i} )+n_{i} (1-n_{j} )\right].                                                                               
\end{equation} 
Here, the bar indicates ensemble average, and $n_{j}$ denotes the mean values of the single particle occupation factors. At zero temperature, occupation factors   are zero or one, and at finite temperatures they are given by the Fermi-Dirac distribution. Single-particle wave functions in each event is determined by the self-consistent mean-field $h[\rho ^{\lambda } (t)]$ of that event according to the TDHF equations,
\begin{equation} \label{eq4} 
i\hbar \frac{\partial }{\partial t} \phi _{j} (\vec{r},t;\lambda )=h[\rho ^{\lambda } (t)]\phi _{j} (\vec{r},t;\lambda ).                                                                                    
\end{equation} 
Simulations of the SMF approach are carried out as follows: In the first step, a complete set of initial wave-functions $\phi _{j} (\vec{r})$ are generated by solving the static Hartee-Fock equations. An event is specified by choosing a set of random matrix elements $\rho _{ij}^{\lambda }$ and boosting the wave functions with proper phase factors. The deformed Hamiltonian at the initial instant $h[\rho ^{\lambda } (t=0)]$ is calculated by the initial density matrix of the event, $\rho ^{\lambda } (\vec{r},\vec{r'},t=0)=\sum _{ij}\phi _{i}^{*}  (\vec{r})\rho _{ij}^{\lambda } \phi _{j} (\vec{r'})$. Then the evolution of the single-particle wave functions $\phi _{j} (\vec{r},t;\lambda )$ is determined by TDHF Eq. \eqref{eq4} while keeping the matrix elements $\rho _{ij}^{\lambda }$ constant. The standard mean-field approximation starting from a deterministic initial condition gives rise to a deterministic final state. On the other hand, the SMF approach generates an ensemble of events where each event is calculated with its own mean-field Hamiltonian. In this approach, it is possible to calculate probability distribution $P_{\lambda } (t)$ of an observable point by point by calculating the expectation value of the observable in each event as,
\begin{equation} \label{eq5} 
Q_{\lambda } (t)=\sum _{ij}\langle\phi _{j} (t;\lambda )|Q| \phi _{i} (t;\lambda )\rangle\rho _{ij}^{\lambda }. 
\end{equation} 
where $Q$ is a one-body operator representing the observable quantity. We should note that, numerical simulations of the SMF can be carried out by employing existing TDHF codes with some modifications \cite{Kim,Umar,Umar2,Sim07,Was08}.

\section{Nucleon Transfer in Deep-Inelastic Collisions}
In order to investigate dynamics of heavy-ion collisions in the framework of the SMF approach, as briefly discussed in the previous section, we need to generate an ensemble of single particle density matrices according to the quantal and thermal density fluctuations in   the initial state. However, in deep-inelastic collisions, since binary character is maintained during the reaction, it is possible to describe gross properties of the reaction mechanism in terms of a set of transport coefficients associated with macroscopic variables. This approach has been very useful in phenomenological nucleon exchange models those developed and applied to analyzed experimental data some years ago \cite{Randrup2,Frei,Adamian}. Therefore, in the initial applications of the SMF, instead of numerical simulations, employing a geometric projection procedure,we extract transport coefficients from the approach. In previous works \cite{Washiyama,Yilmaz,Ayik2}, we carried out this study for central collisions. In this work, we consider more realistic case of off-central collisions of heavy-ions.

\subsection{Window Dynamics}

In deep-inelastic collisions reaction does not lead to fusion. As shown in Fig. 1 binary character of the system is maintained. This figure illustrates the density profile on the reaction plane $\rho (x,y,z=0,t)$ in collision of $^{40}$Ca + $^{40}$Ca  at bombarding energy $E_{\text{cm}}=110$ MeV and initial orbital angular momentum $l=70\hbar$ at three different instants.  Large energy dissipation and angular momentum transfer occur, which are accompanied with large number of nucleon exchange between colliding ions through the window between them. In the figure, symmetry axis ($x'$) and window ($y'$) direction are indicated by red lines. As explained in Appendix A, we can determine the angle between the symmetry axis of the system and the x-axis at each time step by diagonalizing the sigma matrix, which is related to the mass quadrupole tensor of the colliding system. The principal axis of the sigma matrix specifies symmetry axis of the system in the reaction plane ($z=0$) at each time step according to,
\begin{equation} \label{eq6} 
y-y_{\text{cm}} =(x-x_{\text{cm}} )\tan\theta (t) 
\end{equation} 
where ($x_{\text{cm}} $,$y_{\text{cm}} $) denotes coordinates of the center of mass of the system and the rotation $\theta (t)$, given by Eq. \eqref{bsd}, is the angle between the $x$-axis and symmetry axis of the dinuclear shape at time $t$. Window plane is perpendicular to the symmetry axis and should pass through the minimum density location \cite{Was08}. Indicating coordinates of the center of window at the minimum density plane by ($x_{0} $,$y_{0} $), the position of the window plane at time $t$ is represented by,
\begin{equation} \label{eq7} 
y-y_{0} =-(x-x_{0} )\cot\theta (t).     
\end{equation} 
\begin{figure}[hbt]
\includegraphics[width=9cm]{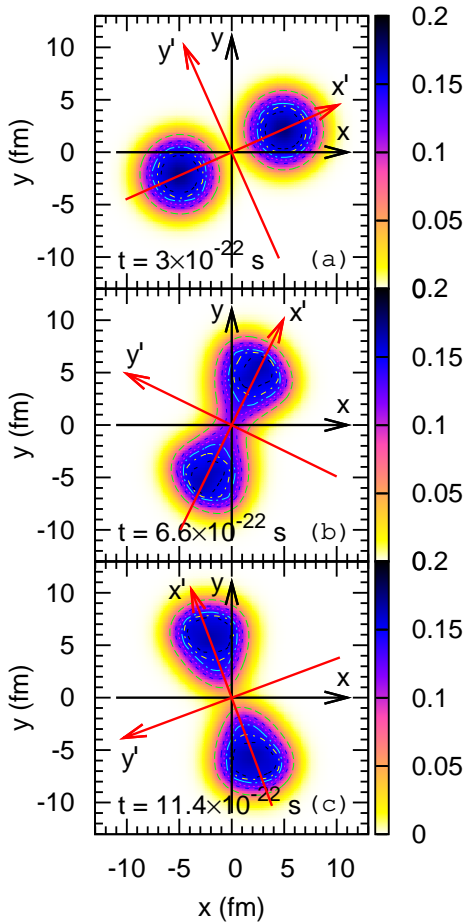}
\caption{(Color online) Snapshots of density profiles on the reaction plane, $\rho (x,y,z=0,t)$, at different times are plotted in units fm$^{-3}$ for $^{40}$Ca + $^{40}$Ca reaction at $E_{\text{cm}}=110$ MeV and initial orbital angular momentum $l=70\hbar$. The panel a), b) and c) correspond respectively to time 
before, during and after contact.}
\end{figure}
For symmetric systems that we are considering in this work, coordinates of the center of mass of the system and the center point of the window are the same, $x_{\text{cm}} =x_{0}$ and $y_{\text{cm}} =y_{0}$. We take the origin of the coordinate system at the center of mass frame, and hence $x_{0} =0$, $y_{0} =0$. However in the following, we present the formulas by keeping the coordinates $(x_{0} ,y_{0} )$ and velocities $(\dot{x}_{0} ,\dot{y}_{0} )$ of the center of the window relative to the center of mass system.

\subsection{Nucleon Diffusion Coefficient}
With the help of the window, we can introduce macroscopic variables associated with binary system, such as mass and charge of target-like and projectile-like fragments, relative position and relative momentum. As a relevant macroscopic variable for the nucleon exchange, we can take the nucleon number of the projectile-like fragments in the event $\lambda$,
\begin{eqnarray} \label{eq8} 
A_{T}^{\lambda } (t)&=&\int \frac{d^{3} p}{\left(2\pi \hbar \right)^{3} }  d^{3} r\Theta\left[(x-x_{0} )\cos \theta +(y-y_{0} )\sin \theta\right]\nonumber\\
&&\quad\times f^{\lambda } (\vec{r},\vec{p},t), 
\end{eqnarray} 
where $\Theta$ is the step function. In each event of symmetric collisions, center position of the window is located at the center of mass, therefore it does not fluctuate. The rotation angle $\theta (t)$ may fluctuate from event to event, which is neglected in this definition. Since it is more convenient to relate the semi-classical approximation, the definition in Eq. \eqref{eq8} is given in terms of phase-space distribution function $f^{\lambda } (\vec{r},\vec{p},t)$. The phase space distribution function is defined as a partial Fourier transform as,
\begin{eqnarray} \label{eq9} 
f^{\lambda } (\vec{r},\vec{p},t)&=&\int d^{3} s\exp \left(-\frac{i}{\hbar } \vec{p}\cdot \vec{s}\right)\nonumber\\
&&\quad\times\rho ^{\lambda } \left(\vec{r}-\frac{\vec{s}}{2} ,\vec{r}+\frac{\vec{s}}{2} ,t\right) 
\end{eqnarray} 
where the density matrix in the event is given by
\begin{equation} \label{eq10} 
\rho ^{\lambda } (\vec{r},\vec{r'},t)=\sum _{ij}\phi _{j}^{*} (\vec{r},t;\lambda )\rho _{ji}^{\lambda }  \phi _{i} (\vec{r'},t;\lambda ). 
\end{equation} 
In the framework of the SMF approach it is possible to deduce Langevin equations for the evolution of the nucleon number of the target-like fragments. For symmetric collisions, since there is no drift, the mean value of the mass asymmetry does not change during the collision. The rate of change of fluctuations in  $A_{T}^{\lambda } (t)$ is determined by the fluctuating part of nucleon flux through the window. In order to calculate the nucleon flux through the window, it is useful to introduce a coordinate transformation to rotating frame in which $x'$ axis is taken along the symmetry axis of the system. Then, we have
\begin{equation} \label{eq11} 
\left(\begin{array}{c} {x'} \\ {y'} \end{array}\right)=\left(\begin{array}{c} {(x-x_{0} )\cos \theta +(y-y_{0} )\sin \theta } \\ {(y-y_{0} )\cos \theta -(x-x_{0} )\sin \theta } \end{array}\right),      
\end{equation} 
and the inverse transformation is given by
\begin{equation} \label{eq12} 
\left(\begin{array}{c} {x-x_{0} } \\ {y-y_{0} } \end{array}\right)=\left(\begin{array}{c} {x'\cos \theta -y'\sin \theta } \\ {y'\cos \theta +x'\sin \theta } \end{array}\right).                                                                                             
\end{equation} 
In the rotating frame, we can express the fluctuating nucleon flux through the window as,
\begin{eqnarray} \label{eq13} 
\frac{\partial }{\partial t} \delta A_{T}^{\lambda } (t)&=&\int \frac{d^{3} p}{(2\pi \hbar )^{3} }  dy'dz'u'_{x} \delta f^{\lambda } (\vec{r},\vec{p},t)|_{x'=0}\nonumber\\
 &=&\xi _{A}^{\lambda } (t). 
\end{eqnarray} 
where, $u'_{x} $ is the component of the velocity along the symmetry axis in the rotating frame and the flux is evaluated across the window plane defined by $x'=0$. The expression of $u'_{x} $ in terms of angular velocity of rotation is given by Eq. \eqref{eq22} in Appendix B.

According to the SMF approach, fluctuating nucleon flux $\xi _{A}^{\lambda } (t)$ across the window acts as a Gaussian random force on the mass-asymmetry variable $A_{T}^{\lambda } (t)$, which is determined by a zero mean value $\overline{\xi _{A}^{\lambda } (t)}=0$ and a second moment $\overline{\xi _{A}^{\lambda } (t)\xi _{A}^{\lambda } (t')}$. In general, second moment involves quantal effects and has a non-Markovian structure. In the present study, we ignore quantal and memory effects and consider transport mechanism in semi-classical approximation. As shown in Appendix B, the expression of the second moment in the semi-classical approximation is given by,
\begin{equation} \label{eq14} 
\overline{\xi _{A}^{\lambda } (t)\xi _{A}^{\lambda } (t')}=2\delta (t-t')D_{AA} (t).                                 
\end{equation} 
where $D_{AA} (t)$ is the diffusion coefficient for nucleon exchange,
\begin{eqnarray} \label{eq15} 
D_{AA} (t)&=&\int \frac{dp_{x} dp_{y} }{(2\pi \hbar )^{2} }  dy'|u'_{x} |\frac{1}{2} \left\{\bar{f}_{T} (t)\left[1-\frac{1}{\Omega } \bar{f}_{P} (t)\right]\right.\nonumber\\&&\qquad+\left.\bar{f}_{P} (t)\left[1-\frac{1}{\Omega } \bar{f}_{T} (t)\right]\right\}. 
\end{eqnarray} 
In this expression, $\bar{f}_{T}(t) =\bar{f}_{T} (x,y,p_{x} ,p_{y} ,t)|_{x'=0} $ and $\bar{f}_{P}(t) =\bar{f}_{P} (x,y,p_{x} ,p_{y} ,t)|_{x'=0} $ represent the reduced Wigner functions on the reaction plane (see Eq. \eqref{eq29} in Appendix B) which are associated with the single particle wave functions originating from target and projectile nuclei, respectively, and $\Omega$ is the volume of the phase space in $(z,p_{z} )$ sub-space (for more detail see \cite{Ayik2}). Since window plane is defined by $x'=0$, the phase-space functions in Eq. \eqref{eq15} depend on the integration variables as, $z=z'$ and
\begin{equation} \label{eq16} 
\left(\begin{array}{c} {x} \\ {y} \end{array}\right)=\left(\begin{array}{c} {x_{0} -y'\sin \theta } \\ {y_{0} +y'\cos \theta } \end{array}\right) .
\end{equation} 
The diffusion coefficient given by Eq. \eqref{eq15} has a similar form as the one familiar from the phenomenological nucleon exchange model. However, since it is deduced from the microscopic SMF approach, it provides more refined description of the transport mechanism. The mean value of drift vanishes for symmetric systems. However, for long interaction times drift can have an effect on diffusion. For sufficiently short interaction times, we can ignore the effect of drift on diffusion. Therefore the width of the fragment mass distribution $\sigma _{AA}^{2} (t)$ is determined by the asymptotic value of the time integral of the diffusion coefficient,
\begin{equation} \label{eq17} 
\sigma _{AA}^{2} (t)=2\int _{0}^{t}dsD_{AA}  (s).                                                                                                         
\end{equation} 

\subsection{Results}
Using the 3D TDHF code \cite{Kim,Sim07,Was08}, we carry out numerical calculations for deep-inelastic collisions of symmetric systems ${}^{40}$Ca + ${}^{40}$Ca and ${}^{90}$Zr + ${}^{90}$Zr. From the time-dependent single-particle wave functions of the occupied states, it is possible to calculate the reduced phase-space distributions $\bar{f}_{T} (x,y,p_{x} ,p_{y} ,t)$ and $\bar{f}_{P} (x,y,p_{x} ,p_{y} ,t)$ originating from target and projectile nuclei. At the separation stage of the reaction, these Wigner functions become very oscillatory, hence it is not possible to directly use these Wigner functions to calculate diffusion coefficient $D_{AA} (t)$ which is derived in the semi-classical approximation. The most important aspect is that the quantal phase-space distributions can become negative in some regions, while in the semi-classical limit phase-space distribution is always positive. In order to obtain the semi-classical approximation, we carry out a smoothing procedure of the quantal phase space distribution (see Appendix C for details). As shown in the example of Fig. 2, the smoothing procedure eliminates negative regions and produces the semi-classical form of the phase-space distribution.
\begin{figure}[htb]
\includegraphics[width=9cm]{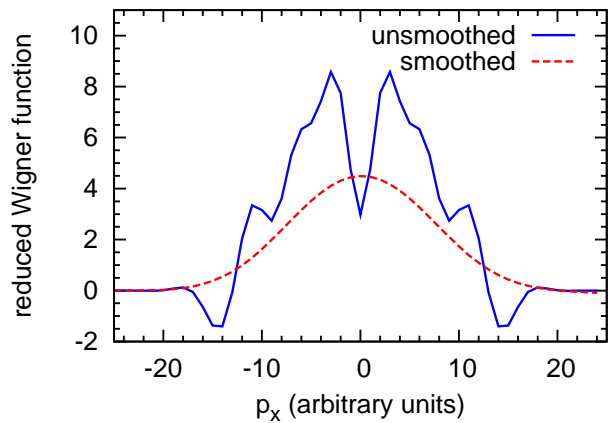}
\caption{(Color online) A plot of the reduced phase-space distribution of one of the fragments $\overline{f}(x=0,y=0,p_x,p_y=0)$ versus $p_x$ at time $t=10^{-21}$ s for ${}^{40}$Ca + ${}^{40}$Ca reaction at $E_{\text{cm}}=110$ MeV and $l=70\,\hbar$. Solid line and dashed line show unsmoothed and smoothed phase-space distributions, respectively.}
\end{figure}
Fig. 3 and Fig. 4 show nucleon diffusion coefficients in collisions of ${}^{40}$Ca + ${}^{40}$Ca and ${}^{90}$Zr + ${}^{90}$Zr systems at bombarding energies 
$E_{\text{cm}}=110$ MeV and $E_{\text{cm}}=300$ MeV, respectively, for different initial orbital angular momenta. The angular momenta 
here is related to the impact parameter $b$ and initial relative velocity $v_{rel}$ through the classical formula $l=\mu\, v_{rel}\,b$, where $\mu$ is the reduced mass. 
\begin{figure}[htb]
\includegraphics[width=9cm]{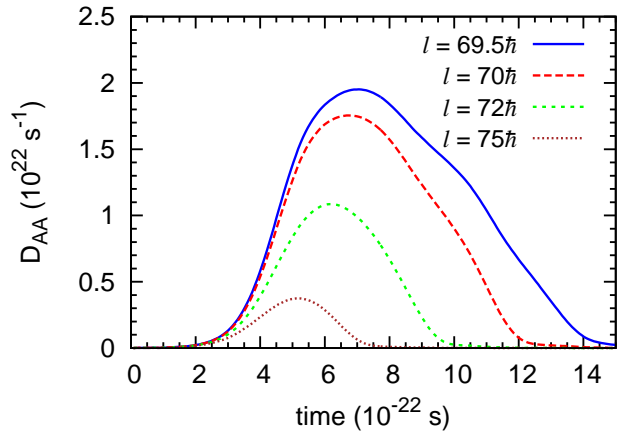}
\caption{(Color online) Diffusion coefficients for nucleon exchange are plotted versus time in ${}^{40}$Ca + ${}^{40}$Ca collisions at bombarding energy $E_{\text{cm}}=110$ MeV and four different initial angular momenta.}
\end{figure}
\begin{figure}[htb]
\includegraphics[width=9cm]{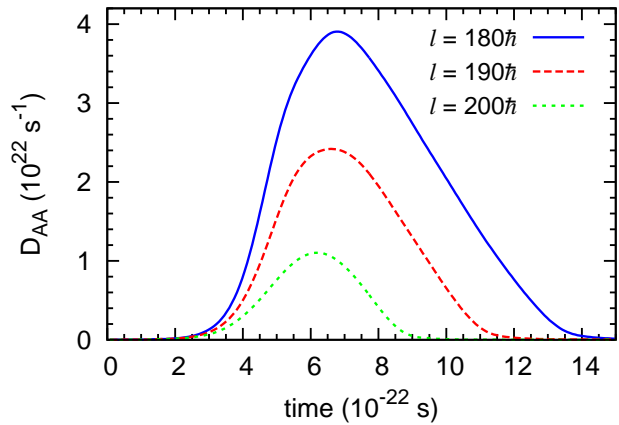}
\caption{(Color online) Diffusion coefficients for nucleon exchange are plotted versus time in ${}^{90}$Zr + ${}^{90}$Zr collisions at bombarding energy $E_{\text{cm}}=300$ MeV and three different initial angular momenta.}
\end{figure}
Fig. 5 and Fig. 6 show the width of the fragment mass distributions for the same systems at the same energies and initial orbital angular momenta as function of time. In these figures, solid lines are the results of calculations by using Eq. \eqref{eq17}, while  symbols indicate results of the empirical formula $\sigma _{AA}^{2} (t)=N_{exc} (t)$. This empirical formula follows from the phenomenological nucleon exchange model and it has been often applied to analyze the experimental data \cite{Frei,Adamian}.
\begin{figure}[htb]
\includegraphics[width=9cm]{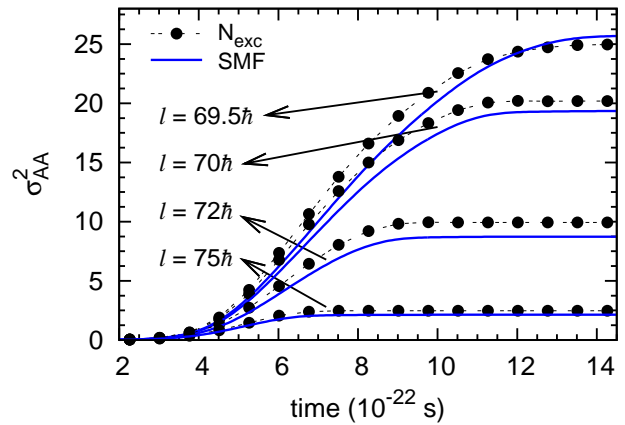}
\caption{(Color online) Widths of fragment mass distributions in collisions of ${}^{40}$Ca + ${}^{40}$Ca  at $E_{\text{cm}}=110$ MeV and four different initial angular momenta. Lines are found by integral of diffusion coefficient, and symbols are results of empirical relation.}
\end{figure}
\begin{figure}[htb]
\includegraphics[width=9cm]{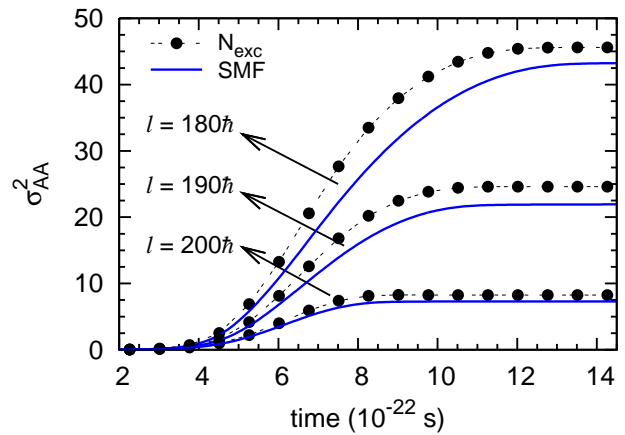}
\caption{(Color online) Widths of fragment mass distributions in collisions of ${}^{90}$Zr + ${}^{90}$Zr  at $E_{\text{cm}}=300$ MeV and three different initial angular momenta. Lines are found by integral of diffusion coefficient, and symbols are results of empirical relation.}
\end{figure}

\begin{figure}[htb]
\includegraphics[width=8cm]{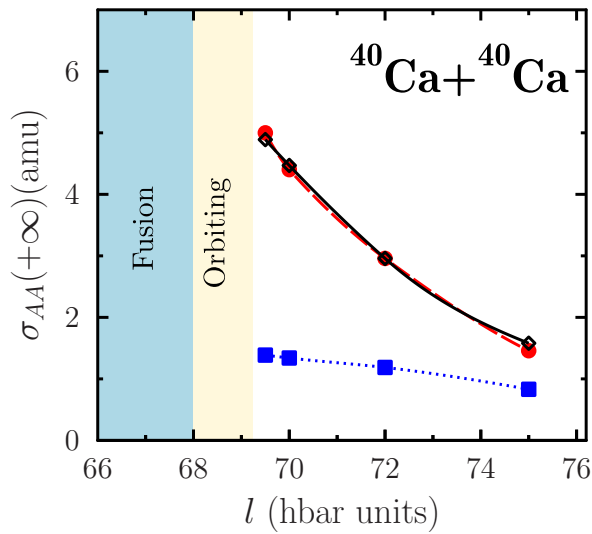}
\caption{(Color online) Asymptotic values of $\sigma_{AA}$ as a
function of the orbital angular momentum obtained
with SMF (red filled circles) and TDHF (blue filled squares). The asymptotic value $\sqrt{N_{\rm exc} (\infty)}$ is also shown by black open squares.
The blue area indicates the Fusion region while the yellow area indicates the "Orbiting" region where the two 
nuclei stick together for a long time and then re-separate.}
\label{fig:sigmaAinf}
\end{figure}

The fact that the results of SMF calculations are consistent with the empirical formula provides a strong support for the validity of the SMF approach. 
Fig. 7 shows the asymptotic values $\sigma_{AA}(\infty)$ of the SMF dispersions of the fragment mass distributions in  ${}^{40}$Ca + ${}^{40}$Ca  collisions at bombarding energy $E_{\text{cm}}=110$ MeV for four different impact parameters. We note that since we neglect the drift term in the SMF description, our calculations are not valid for long interaction times, like in the case of the orbiting processes, during which the fragment mass distribution may reach the
equilibrium limit.  For this reason, we only consider $L>69.5 \hbar$, that corresponds to those initial angular momenta $L$ outside the orbiting region 
(see yellow area in Fig. \ref{fig:sigmaAinf}). 
There is data available for  ${}^{40}$Ca + ${}^{40}$Ca  at a slightly higher bombarding energy $E_{\text{cm}}=128$ MeV and over a different angular range
\cite{Roynette}. Dispersion of the fragment mass distributions over the measured angular range is found between $\sigma_{AA}=2.8$ amu and $4.6$ amu. 
Results of the SMF calculations are consistent with these measurements.

In the standard mean-field approximation, dispersion of the fragment mass distribution is determined by,
\begin{equation} \label{eq19} 
\sigma _{AA}^{2} (t)=\sum _{ij}|\langle\phi _{j} (t)|\Theta |\phi _{i} (t)\rangle|^{2}  n_{i} (1-n_{j} ) 
\end{equation} 
where $\Theta =\Theta [(x-x_{0} )\cos \theta +(y-y_{0} )\sin \theta ]$. Using the completeness relation and the relation, $\Theta ^{2} =\Theta $, this expression can be evaluated in terms of occupied, i.e. hall states, as follows,
\begin{eqnarray} \label{eq20} 
\sigma _{AA}^{2} (t)&=&\sum _{h}\langle\phi _{h} (t)|\Theta |\phi _{h} (t)\rangle|\nonumber\\
 &&-\sum _{h,h'}|\langle\phi _{h} (t)|\Theta |\phi _{h'} (t)\rangle|^{2}.   
\end{eqnarray} 
Solid squares in Fig. 7 indicate the asymptotic values of dispersion of fragment mass distributions in the collisions of ${}^{40}$Ca + ${}^{40}$Ca at the same bombarding  energies and the same initial orbital angular momenta. We observe that, in particular for long interaction times associated with large energy loss, the standard mean-field approximation severely underestimates dispersions of the fragment mass distributions.

\section{Conclusions}

The SMF approach goes beyond the standard mean-field approximation by including quantal and thermal fluctuations in the initial state. This approach constitutes an approximate treatment of the effect of superposition of many Slater determinants. As a result the approach provides a description of fluctuations of the collective motion at sufficiently low energies at which collisional dissipation does not play an important role. The standard mean-filed approximation is deterministic in the sense that a well defined initial condition leads to a  unique final state.  On the other hand, in the SMF starting from a well-defined distribution of the density fluctuations in the initial state, an ensemble of single-particle density matrices are generated by evolving each event by its own self-consistent mean-field. It is possible to generate distribution function of an observable by calculating the expectation values of the observable event by event. In the present work, rather than carrying out stochastic simulations, we extract diffusion coefficient for nucleon exchange in deep-inelastic heavy-ion collisions by employing a projection procedure. By extending a previous study, we investigate nucleon exchange mechanism in off-central collisions and calculate nucleon diffusion coefficients for collisions of symmetric ${}^{40}$Ca + ${}^{40}$Ca and ${}^{90}$Zr + ${}^{90}$Zr systems. The width of fragment mass distributions calculated in the framework of the SMF approach are consistent with the empirical formula often employed to analyze experimental data. There is experimental data for the dispersion of the fragment mass distribution of ${}^{40}$Ca + ${}^{40}$Ca collisions at bombarding energy $E_{\text{cm}}=128$. Our calculations give a reasonable magnitude of the mass dispersion compared to the measured values, 
while the standard mean-field approximation predicts much smaller values in particular in collisions with large energy dissipation. Work is in progress to extend the present approach to asymmetric systems for off-central collisions. In particular, it was recently shown that pairing alone cannot explain the large enhancement of two-particle transfer channel 
observed experimentally in $^{40}$Ca + $^{96}$Zr \cite{Sca13,Cor11}.  The inclusion of quantal zero point motion in collective space through the SMF technique is anticipated to improve the agreement between experiment and microscopic mean-field theories.
 
\begin{acknowledgments}
S.A. gratefully acknowledges TUBITAK and IN2P3-CNRS, Universit\'e Paris-Sud for partial support and warm hospitality extended to him during his visits. This work is supported in part by US DOE Grant No. DE-FG05-89ER40530, and in part by TUBITAK Grant No. 113F061. We also thank K. Washiyama for discussions at the early stage of this work.
\end{acknowledgments}

\appendix
\section{}
The symmetry axis of dinuclear shape can be easily obtained when the reaction plane is $z=0$. The position variances can be used to form a matrix as
\begin{equation}
\sigma(t)=\left( \begin{array}{cc}
\sigma_{xx}(t) & \sigma_{xy}(t) \\
\sigma_{xy}(t) & \sigma_{yy}(t) \end{array} \right),
\end{equation}
where 
\begin{equation}
 \sigma_{ab}(t)=\sum_{j=1}^A\langle \phi_j| x_{a}x_{b}|\phi_j\rangle-\sum_{i,j=1}^A\langle\phi_i| x_{a}|\phi_i\rangle\langle\phi_j| x_{b}|\phi_j\rangle,
\end{equation}
with $a,b=x,y$. The eigenvectors of the matrix $\sigma(t)$ are 
\begin{equation}
\vec{\sigma}_{\pm}=\left(\begin{array}{c}
\sigma_{\pm}(t)\\
1 \end{array}\right),
\end{equation}
where the components of the eigenvectors are given by,
\begin{equation}
 \sigma_{\pm}(t)=\frac{\sigma_{xx}-\sigma_{yy}\pm\sqrt{(\sigma_{xx}-\sigma_{yy})^2+4\sigma^2_{xy}}}{2\sigma_{xy}}.
\end{equation}
The components of the eigenvectors of the matrix $\sigma(t)$ form the $x$ and $y$ coordinates of the transformed axes $x'$ and $y'$. Hence, $\vec{x'}$ is the vector from the origin $(0,0)$ to the point $(\sigma_+,1)$ and $\vec{y'}$ is the vector from the origin to the point $(\sigma_-,1)$ in the fixed $x,y$ coordinate system. 
Then, the angle between $x$ axis and $x'$ axis (symmetry axis) is
\begin{eqnarray}
 \label{asd}
\tan\theta&=&\frac{1}{\sigma_+}\nonumber\\
&=&\frac{2\sigma_{xy}}{\sigma_{xx}-\sigma_{yy}+\sqrt{(\sigma_{xx}-\sigma_{yy})^2+4\sigma^2_{xy}}}.
\end{eqnarray}
Comparing the following trigonometric transformation,
\begin{equation}
 \tan\theta = \frac{\tan2\theta}{1 + \sqrt{1+\tan^2 2\theta}}, \quad \theta \in \left(-\frac{\pi}{2},\frac{\pi}{2} \right),
\end{equation}
with Eq. \eqref{asd}, the angle can be written as
\begin{eqnarray}\label{bsd}
 \theta=\frac{1}{2}\tan^{-1}\left(\frac{2\sigma_{xy}}{\sigma_{xx}-\sigma_{yy}}\right).
\end{eqnarray}

\section{}
We consider correlation function of the total phase-space distribution $f^{\lambda } (\vec{r},\vec{p},t)$ in the event $\lambda $. During a short time interval, we can approximately express time evolution as free propagation,
\begin{equation} \label{csd} 
\delta f^{\lambda } (\vec{r},\vec{p},t+\tau )=\delta f^{\lambda } (\vec{r}-\vec{u}\tau ,\vec{p},t) 
\end{equation} 
where $\vec{u}=\vec{v}-\dot{\vec{\theta }}\times (\vec{r}-\vec{r}_{0} )$ denotes the velocity relative to window. Here $\dot{\vec{r}}_{0} $ and $\vec{r}_{0} $ are translational velocity and position of the center of the window relative to the center of mass of the total system, and $\dot{\vec{\theta }}$ denotes the rotational velocity of the window about $z$-axis in the reaction plane. Following the discussion presented in the Appendix of ref. \cite{Ayik2}, we can write the correlation function of phase-space fluctuations as, 
\begin{eqnarray} \label{eq21} 
&&\overline{\delta f^{\lambda } (\vec{r}_{1} ,\vec{p}_{1} ,t)\delta f^{\lambda } (\vec{r}_{2} ,\vec{p}_{2} ,t')}=(2\pi \hbar )^{3} \delta (\vec{p}_{1} -\vec{p}_{2} )\nonumber\\
&&\qquad\qquad\times\delta \left[\vec{r}_{1} -\vec{r}_{2} -(t-t')\vec{u}\right]\Lambda ^{+} (\vec{r}_{1} ,\vec{p}_{1} ,t) 
\end{eqnarray} 
where
\begin{eqnarray} \label{eq22} 
\Lambda ^{+} (\vec{r},\vec{p},t)&=&\left\{f_{P} (\vec{r},\vec{p},t)\left[1-f_{T} (\vec{r},\vec{p},t)\right]\frac{}{}\right.\nonumber\\
&&\left.\frac{}{}+f_{T} (\vec{r},\vec{p},t)\left[1-f_{P} (\vec{r},\vec{p},t)\right]\right\}. 
\end{eqnarray} 
Here, $f_{P} (\vec{r},\vec{p},t)$ and $f_{T} (\vec{r},\vec{p},t)$ are phase-space distributions associated with wave functions originating from projectile and target, respectively. We can decompose the delta function in the center of mass frame as,
\begin{eqnarray} \label{eq23} 
&&\delta \left[\vec{r}_{1} -\vec{r}_{2} -(t-t')\vec{u}\right]=\delta \left[x_{1} -x_{2} -(t-t')u_{x} \right]\nonumber\\
&&\times\delta \left[y_{1} -y_{2} -(t-t')u_{y} \right]\delta \left[z_{1} -z_{2} -(t-t')u_{z} \right] 
\end{eqnarray} 
where $x$-axis denotes the beam direction. In the center of mass frame, the components of velocity are $u_{x} =v_{x}$ and 
\begin{equation} \label{eq24} 
\left(\begin{array}{c} {u_{x} } \\ {u_{y} } \end{array}\right)=\left(\begin{array}{c} {v_{x} -\dot{x}_{0} +\dot{\theta }(y-y_{0} )} \\ {v_{y} -\dot{y}_{0} -\dot{\theta }(x-x_{0} )} \end{array}\right).                                                                                                      
\end{equation} 
It is also possible to decompose the delta function in the rotating frame as,
\begin{eqnarray} \label{eq25} 
&&\delta [\vec{r}_{1} -\vec{r}_{2} -(t-t')\vec{u}]=\delta [x'_{1} -x'_{2} -(t-t')u'_{x} ]\nonumber\\
&&\times\delta [y'_{1} -y'_{2} -(t-t')u'_{y} ]\delta [z'_{1} -z'_{2} -(t-t')u'_{z} ].                  
\end{eqnarray} 
Components of the velocity in the rotating frame relative to the center of the window are given by $u'_{x} =v_{x}$ and
\begin{eqnarray} \label{eq26} 
&&\left(\begin{array}{c} {u'_{x} } \\ {u'_{y} } \end{array}\right)=\left(\begin{array}{c} {u_{x} \cos \theta +u_{y} \sin \theta } \\ {u_{y} \cos \theta -u_{x} \sin \theta } \end{array}\right)\nonumber\\
&&=\left(\begin{array}{c} {(v_{x} -\dot{x}_{0} )\cos \theta +(v_{y} -\dot{y}_{0} )\sin \theta +\dot{\theta }y'} \\ {(v_{y} -\dot{y}_{0} )\cos \theta -(v_{x} -\dot{x}_{0} )\sin \theta -\dot{\theta }x'} \end{array}\right). 
\end{eqnarray} 
It is more convenient to express the correlation function Eq. \eqref{eq17} of the phase-space distribution in the rotating frame. Since on the window $x'_{1} =x'_{2} =0$, the correlation function can be given as,
\begin{eqnarray} \label{eq27} 
&&\overline{\delta f^{\lambda } (\vec{r}_{1} ,\vec{p}_{1} ,t)\delta f^{\lambda } (\vec{r}_{2} ,\vec{p}_{2} ,t')}=(2\pi \hbar )^{3} \delta (\vec{p}_{1} -\vec{p}_{2} )\nonumber\\
&&\times\frac{\delta (t-t')\delta (y'_{1} -y'_{2} )\delta (z'_{1} -z'_{2} )}{|u'_{x} |} \Lambda ^{+} (\vec{r},\vec{p},t)|_{x'=0}.  
\end{eqnarray} 
Using this result we can calculate the nucleon diffusion coefficient to obtain the expression given by Eq. \eqref{eq15}. In order to simplify the numerical calculations of the diffusion coefficient, in the integrations perpendicular to the reaction plane in Eq. \eqref{eq15}, we introduce the approximation,
\begin{eqnarray} \label{eq28} 
\int \frac{dp_{z} dz}{(2\pi \hbar )^{2} }  f_{P} (\vec{r},\vec{p},t)f_{T} (\vec{r},\vec{p},t)&=&\frac{1}{\Omega } \bar{f}_{P} (x,y,p_{x} ,p_{y} ,t)\nonumber\\
&&\bar{f}_{T} (x,y,p_{x} ,p_{y} ,t),                                      
\end{eqnarray} 
where $\Omega $ is the volume of the reduced phase space $(z,p_{z} )$ and  
\begin{equation} \label{eq29} 
\bar{f}_{P/T} (x,y,p_{x} ,p_{y} ,t)=\int \frac{dp_{z} dz}{(2\pi \hbar )^{2} }  f_{P/T} (\vec{r},\vec{p},t) 
\end{equation} 
denotes the reduced phase-space distributions on the reaction plane.

\section{}
The smoothing of the reduced Wigner functions is performed at each time step by using the NAG library subroutine G10ABF which does cubic spline smoothing. G10ABF is smoothing in one dimension. In order to obtain two dimensional ($p_x,p_y$) smoothing first smoothing is performed over one of the dimensions and then over the other dimension. Various smoothing parameters has been tried and it is concluded that a smoothing parameter value 500 for both dimensions is close to optimum. 

The unsmoothed and smoothed reduced Wigner functions are tested by taking their integrals over momenta which give the local reduced density of one of the fragments as
\begin{equation}
 \bar{\rho}_T(x,y,t)= \int \frac{p_xp_y}{(2\pi\hbar)^2}\,\bar{f}_{T} (x,y,p_{x} ,p_{y} ,t),
\end{equation}
where $\bar{\rho}_T(x,y,t)=\int dz\, \rho_T(x,y,z,t)$. It is seen that the reduced density is obtained for unsmoothed as well as smoothed reduced Wigner functions.

\end{document}